\documentclass[usenatbib]{mnras}

\usepackage{mathptmx}

\usepackage[T1]{fontenc}
\usepackage{ae,aecompl}

\usepackage{amsmath}	
\usepackage{amssymb}	
\usepackage{graphicx}	

\title[SIV, MOND and Dwarf Spheroidals]{Scale-Invariant Dynamics of Galaxies, \\ MOND, Dark Matter, and the Dwarf Spheroidals}
\author[Maeder \& Gueorguiev]{
Andre Maeder $^{1}$\thanks{E-mail: andre.maeder at unige.ch}
and Vesselin G. Gueorguiev $^{2,3}$\thanks{E-mail: vesselin.gueorguiev at ronininstitute.org}\\
$^{1}${Geneva Observatory - chemin des Maillettes 51, CH-1290 Sauverny, Switzerland}\\
$^{2}${Institute for Advanced Physical Studies - 21 Montevideo Street, Sofia 1618, Bulgaria}\\
$^{3}${Ronin Institute for Independent Scholarship - 127 Haddon Pl., Montclair, NJ 07043, USA}
}

\date{Accepted XXX. Received YYY; in original form ZZZ}

\pubyear{2019}

\begin{document}
\label{firstpage}
\pagerange{\pageref{firstpage}--\pageref{lastpage}}
\maketitle

\begin{abstract}
The  Scale-Invariant Vacuum (SIV)  theory is based on Weyl's Integrable Geometry, 
endowed with a gauge scalar field. The main difference between MOND and the SIV theory is that the first
considers a global dilatation invariance of space and time, where the scale factor $\lambda$
is a constant, while the second opens the likely possibility that $\lambda$ is a function of time.
The key equations of the SIV framework are used here to study the  relationship between the Newtonian
gravitational acceleration due to baryonic matter $g_{\mathrm{bar}}$
and the observed kinematical acceleration $g_{\mathrm{obs}}$. 
The relationship is applied to galactic systems of the same age where
the Radial Acceleration Relation (RAR), between the $g_{\mathrm{obs}}$ and
$g_{\mathrm{bar}}$ accelerations, can be compared with observational data.
The SIV theory shows an excellent agreement with observations and with MOND
for baryonic gravities $g_{\mathrm{bar}}>10^{-11.5}$ m s$^{-2}$.
Below this value, SIV still fully agrees with the observations, as
well as with the horizontal asymptote of the RAR for dwarf spheroidals,  
while this is not the case for MOND. These results support the view   that there is no need for 
dark matter and that the RAR and related dynamical properties of galaxies 
can be interpreted by a modification of gravitation.
\end{abstract}




\section{Introduction}

The Scale-Invariant Vacuum (SIV) theory is based on the Weyl Integrable
Geometry. A general scale-invariant field equation and a geodesic
equation have been obtained first by \citet{Dirac73} and \citet{Canu77}
and their consistency as a framework for gravitation and the motion
of astronomical bodies have been further developed by \cite{BouvierM78}.


``It appears as one of the fundamental principles in Nature that
the equations expressing basic laws should be invariant under the
widest possible group of transformations'' \citep{Dirac73}. The
scale or gauge invariance of physical laws is a fundamental property
in physics. Maxwell's equations are scale-invariant in the empty space,
this is also true for General Relativity (GR) if the cosmological
constant $\Lambda_{\mathrm{E}}$ is absent. Scale invariance
means that the equations do not change for a transformation of the
line element of the form, 
\begin{equation}
ds'\,=\,\lambda(t)\,ds\,,\label{ds}
\end{equation}
where $ds'$ is the line element of GR and $ds$ the line element of
a more general space where scale invariance is also present. 
The term $\lambda(t)$ is the scale factor. It is considered to not depend on space
for reason of homogeneity and isotropy. The requirement of scale invariance
in addition to the general covariance implies to move from the Riemann
Geometry to Weyl's Geometry, which in addition to a metric form 
$ds^2 = g_{\mu \nu} dx^{\mu} dx^{\nu}$
is endowed with a scalar field $\Phi$ (see Equation \ref{k2} below).
By adding the gauging condition that the macroscopic empty space is scale-invariant, 
the corresponding general scale-invariant field and geodesic equations obtained by \citet{Canu77}
have been applied successfully to several basic cosmological tests \citep{Maeder17a},
to clusters of galaxies and galactic properties \citep{Maeder17c},
as well as to the growth of density fluctuations in the early Universe
\citep{MaedGueor19}.


The astrophysical problem considered here, within the scale-invariant
framework as well as in the context of MOND, concerns the Radial
Acceleration Relation (RAR) of galaxies. The RAR compares the centripetal
acceleration $g_{\mathrm{obs}}$, traced by the rotation curves of
spiral galaxies, and the expected gravitational acceleration 
due to the observed distribution of baryons $g_{\mathrm{bar}}$ \citep{McGaugh04,McGaugh16,Lelli17,Li18}.
Below a gravity of about $10^{-10}$ m s$^{-2}$, the RAR deviates
from the 1:1 line, $g_{\mathrm{obs}}$ being much larger than $g_{\mathrm{bar}}$.
The RAR is followed by late and early type galaxies, and also by the
dwarf spheroidals where the deviations from the 1:1 line are the largest
ones.

These deviations are currently attributed to dark matter. \citet{Lelli17}
point out that the dark matter distribution is fully determined by
that of the baryons or vice-versa. A number of authors have interpreted
this relation in the context of the $\Lambda$CDM models of galaxy
formation, in terms of different mass-dependent density profiles
of DM haloes \citep{Dicintio16}, of simulations of galaxy formation
matching the velocities and the scaling relations \citep{Santos16},
in particular if stellar masses and sizes are closely related to the
masses and sizes of their DM haloes \citep{Navarro17,Desmond17}.
\citet{Keller17} show that the account for the hot outflows from
supernovae at high $z$ improves the comparisons by producing a baryon
depletion, while the effects of the AGN feedback is considered by
\citet{Ludlow17}.

Attempts to explain the RAR in the context of modified gravity (MOND)
have been successfully made by \citet{Milgrom16} and \citet{Li18}.
The MOND theory \citep{Milgrom83,Milgrom09} shows some significant
successes in explaining galactic properties, such as the flat rotation
curves of spiral galaxies, the radial acceleration relation (RAR)
of galaxies \citep{McGaugh16,Lelli17}, and the Tully-Fisher relation.
Its mutual advantages and disadvantages with respect to the current
$\Lambda$CDM model have been analyzed by \citet{McGaugh15}. The
MOND theory also presents some properties of coordinate scale invariance
\citep{Milgrom09,Milgrom14}, such that 
\begin{equation}
(t,{\bf {r}})\,\rightarrow\,(\lambda\,t,\lambda{\bf {r})\,,}
\end{equation}
where here $\lambda$ is a constant term. This is a global dilatation
invariance independent on time and space, much less constraining than
a scale invariance to a transformation like in equation (\ref{ds}).
Thus, the main difference between MOND and SIV theory is that the first
does not consider a possible time variation of the scale factor $\lambda$, 
while the second allows this possibility. 
This makes a big difference because the first and second derivatives of 
$\lambda$ will appear in the dynamical equations.
In this context, it is interesting to investigate the possible relation, if any, 
between MOND and the SIV framework \citep{Maeder17a}
that rests on a more general time-dependent scale invariance.

In this work, new research lines are explored, a justified objective,
especially in the present context, where the dominant mass-energy source is unknown.
In Section \ref{base}, we summarize the main relevant properties
of the Integrable Weyl Geometry which serves as a basis for further
developments. In Section \ref{gauge_condition}, we express the field
and geodesic equations, as well as the adopted gauge condition. Section
\ref{acceleration_relation} applies the weak-field equation and derives
the relation between the kinematic rotational acceleration and the
baryonic matter present. Section \ref{Mobs} presents the comparison
of the results with observations and compares to the MOND predictions.
Section \ref{Concl} is devoted to the conclusion summary.

\section{Key Equations of the Integrable Weyl's Geometry}

\label{base}

Weyl's Geometry is a generalization of the Riemann Geometry first
proposed by Hermann Weyl \citep{Weyl23} and further developed by
\citet{Eddington23} and \citet{Dirac73}. 
Weyl's Geometry is the appropriate framework 
to study scale invariance problems, in particular 
the corresponding Integrable Weyl's Geometry \citet{Canu77} and \citet{BouvierM78}.
In addition to the general
covariance of GR, it considers gauge or scale  transformations.
The original aim of Weyl was to interpret the electromagnetism in
terms of properties of the space-time geometry, as Einstein did for
gravitation. This geometry is endowed with a metrical determination
of the quadratic form $ds^{2}\,=\,g_{\mu\nu}(x)dx^{\mu}dx^{\nu}$
\, as in Riemann space, and with quantities expressing gauge transformations.
Let us consider a vector of length $\ell$ attached at a point $P$
of coordinates $x^{\mu}$. If this vector is transported by parallel
displacement to a point $P'$ of coordinates $x^{\mu}+\delta x^{\mu}$,
its length becomes $\ell+\delta\ell$, where 
\begin{equation}
\delta\ell\,=\,\ell\kappa_{\mu}\delta x^{\mu}\,,\label{k1}
\end{equation}
where $\kappa_{\mu}$ is called the coefficient of metrical connection.
In Weyl's geometry, terms $\kappa_{\mu}$ are fundamental coefficients
as are the $g_{\mu\nu}$ in GR. The lengths
undergo corresponding gauge changes: 
\begin{equation}
\ell'\,=\,\lambda(x)\,\ell\,,\label{l1}
\end{equation}
where $\lambda$  is the scale gauge factor,  which could in principle depend on the 4--coordinates. 
  The segment $\delta\ell'$ also changes. To the first order
in $\delta x^{\nu}$, one has: 
\begin{equation}
\ell'+\delta\ell'\,  =  \,(\ell+\delta\ell)\lambda(x+\delta x)=(\ell+\delta\ell)\lambda(x)+\ell\frac{\partial\lambda}{\partial x^{\nu}}\delta x^{\nu}\,,
\end{equation}
\begin{eqnarray}
\delta\ell'\,  &=& \,\lambda\delta\ell+\ell\lambda_{,\nu}\delta x^{\nu}=
\lambda\ell\kappa_{\nu}\delta x^{\nu}+\ell\lambda_{,\nu}\delta x^{\nu} \,= \\ \nonumber
&=&  \lambda\ell\left(\kappa_{\nu}+\Phi_{,\nu}\right)\delta x^{\nu}\,=l'\kappa'_{\nu}\delta x^{\nu}\,,
\end{eqnarray}
where the notation $\lambda_{,\nu}=\frac{\partial\lambda}{\partial x^{\nu}}$
is used along with: 
\begin{equation}
\Phi\,=\ \ln\lambda\,,\quad\quad\mathrm{and}\quad\kappa'_{\nu}=\kappa_{\nu}+\Phi_{,\nu}=\kappa_{\nu}+\partial_{\nu}\ln\lambda\,.
\label{k2}
\end{equation}
If the vector is parallel transported along a closed loop, the total
change of the length of the vector can be written as: 

\begin{equation}
\Delta\ell\,=\,\ell\left(\partial_{\nu}\kappa_{\mu}-\partial_{\mu}\kappa_{\nu}\right)\,\sigma^{\mu\nu}\,,
\label{boucle}
\end{equation}
where 
$\sigma^{\mu\nu}=dx^{\mu}\wedge dx^{\nu}$ 
is an infinitesimal surface element corresponding to the edges 
$dx^{\mu}$ and $dx^{\nu}$. The tensor 
$F_{\mu\nu}\,=\,\left(\kappa_{\mu,\nu}-\kappa_{\nu,\mu}\right)$
was identified by Weyl with the electromagnetic field. However, in
the above form of the Weyl's geometry, the lengths are non-integrable:
the change of the length of a vector between two points depends on
the path considered. Such a property would imply that different atoms,
due to their different world lines, would have different properties
and thus will emit at different frequencies. This was the essence
of Einstein's objection against Weyl's geometry, as recalled by \citet{Canu77}.
\\

\noindent 
The above objection does not hold if one considers the so-called
Integrable Weyl's Geometry, which forms a consistent framework for
the study of gravitation as emphasized by \citet{Canu77} and \citet{BouvierM78}.
Let us consider that the framework of functions denoted by primes
is the Riemann space as in GR. That is, $\kappa'_{\nu}=0$, and therefore:

\begin{equation}
\kappa_{\nu}\,=\,-\Phi_{,\nu}\,=\,-\frac{\partial\ln\lambda}{\partial x^{\nu}}\,.\label{k3}
\end{equation}
The above condition means that the metrical connection $\kappa_{\nu}$
is the gradient of a scalar field $(\Phi=\ln\lambda)$, as seen above.
Thus, $\kappa_{\nu}dx^{\nu}$ is an exact differential and therefore:
\begin{equation}
\partial_{\nu}\kappa_{\mu}\,=\,\partial_{\mu}\kappa_{\nu}\,,\label{dx}
\end{equation}
which according to Equation (\ref{boucle}) implies that the parallel
displacement of a vector along a closed loop does not change its length,
equivalently the change of the length due to a displacement does not
depend on the path followed. Many mathematical tools of Weyl's geometry
also work in the integrable form of this geometry. The line element
$ds'=g'_{\mu\nu}dx^{\mu}dx^{\nu}$ refers to GR, while $ds$ in 
Equation (\ref{ds}) refers to the Integrable Weyl Geometry, which 
is also endowed with a scalar gauge field $\Phi $.\\

\noindent 
The integrable Weyl's space is conformally equivalent to
a Riemann space (pseudo-space) defined by $g'_{\mu\nu}$ with $\kappa'_{\nu}=0$
via the $\lambda$ mapping which represents gauge re-scaling: 
\begin{equation}
g'_{\mu\nu}\,=\,\lambda^{2}\,g_{\mu\nu}\,.\label{conformal}
\end{equation}
\noindent
Such conformal mappings have been related to space-time deformations
and have been studied in connection with a large class of Extended
Theories of Gravity \citep{Capozziello07,Capozziello11}. 
When the traceless symmetric and antisymmetric components of these
general transformations are absent the corresponding generalized conformal
transformations become simple conformal mappings. In this respect, Eq.
(\ref{conformal}) relates the original unmodified metrics and the
``deformed metrics'' of the scale-invariant vacuum space-time. According
to \citet{Capozziello07}, the space-time deformations act like a
force that deviates the test particles from the unperturbed motions
and these authors also express the corresponding geodesics and general
field equation that are consistent with \citet{Canu77} and \citet{BouvierM78}.\\


\noindent 
The conformal factor that is only time dependent is an important ingredient
in  the  SIV theory since it allows space-time deformations that obey the
cosmological principle and keep the space homogenous and isotropic
at every moment of time. If $\lambda$ is to depend on
the space coordinates, i.e, as in Brans-Dicke theory of gravity, then
one would face such undesirable inhomogeneity problems. Furthermore,
due to the Universality of the Einstein theory of gravitation \citep{Kijowski2016},
any Extended Theories of Gravity can be put into an equivalent standard
GR theory with (possibly) a different metric tensor plus (possibly)
a different set of matter fields. There could be at least one effective
scalar field as in Brans-Dicke theory. Here, this scalar field is
that expressed by the metrical connection $\kappa_{\nu}$. Furthermore,
we may remark that in the scale-invariant vacuum theory the ``deformation''
is not necessarily a perturbation of the standard motion, it can even
become the dominant effect as illustrated by the scale-invariant cosmological
models \citep{Maeder17a} which show that the accelerated expansion
becomes the leading term in the advanced evolutionary stages of the Universe.  
An  interesting case of Extended Theories of Gravity  is the case of the ``$f(R)$- theory of Gravity'' 
\citep{Capozziello06}, where the Lagrangian contains some function  $f(R)$ of the Ricci curvature scalar $R$
(instead of just $R$). Such theories have been developed and also applied to the dynamics of spiral and 
elliptical galaxies by \citet{Capozziello17} and will be further discussed in Section \ref{Comp}.

In the framework of Weyl's Geometry, scalars, vectors, or tensors
that transform like 
\begin{equation}
Y'^{\, \nu   }_{\mu}  \, =  \, \lambda^{n} \, Y^{\nu}_{\mu} \, ,
\label{co}
\end{equation}
are respectively called co-scalars, co-vectors, or co-tensors of power
$n$. If $n=0$, one has an in-scalar, in-vector or in-tensor, such
objects are invariant upon a scale transformation. Scale covariance
refers to a transformation with powers $n$ different from zero, while
the term scale invariance is generally reserved for cases with $n=0$. On the
basis of the properties recalled above, a so-called cotensor analysis
has been developed \citep{Weyl23,Eddington23,Dirac73, Canu77}. 
\citet{BouvierM78} have also derived the equation of geodesics from an action principle
and shown its consistency with the notion of the shortest distance between
two points, they have also reviewed  the notion of parallel displacement,
of isometries and Killing vectors in the Integrable Weyl's Geometry.
In general, the derivative of a scale-invariant object is not scale
invariant. Thus, scale covariant derivatives of the first and second
order have been developed preserving scale covariance. For example,
the co-covariant derivatives $A_{\mu*\nu}$ and $A_{*\nu}^{\mu}$
of a co-vector $A_{\mu}$ of power n are: 
\begin{eqnarray}
A_{\mu*\nu}\, & = & \,\partial_{\nu}A_{\mu}-^{*}\Gamma_{\mu\nu}^{\alpha}A_{\alpha}-n\kappa_{\nu}A_{\mu}\,,\\{}
A_{*\nu}^{\mu}\, & = & \,\partial_{\nu}A^{\mu}+^{*}\Gamma_{\nu\alpha}^{\mu}A^{\alpha}-n\kappa_{\nu}A^{\mu},\,\label{eq:co-cov_der}\\
\mathrm{with}\quad^{*}\Gamma_{\mu\nu}^{\alpha} & = & \Gamma_{\mu\nu}^{\alpha}+g_{\mu\nu}\kappa^{\alpha}\,-g_{\mu}^{\alpha}\kappa_{\nu}-g_{\nu}^{\alpha}\kappa_{\mu}.\label{eq:co-Chr_sym}
\end{eqnarray}
Here $^{*}\Gamma_{\mu\nu}^{\alpha}$ is a modified Christoffel symbol,
while $\Gamma_{\mu\nu}^{\alpha}$ is the usual Christoffel symbol.
For more details on the cotensor calculus, the interested reader may
read Chapter VII of ``The Mathematical Theory of Relativity \citep{Eddington23},
as well as \citet{Dirac73}. We also point out that \citet{Canu77}
provided a short summary of the cotensor analysis. In this framework,
the Riemann curvature tensor $R_{\mu\lambda\rho}^{\nu}$, its contracted
form, the Ricci tensor $R_{\mu}^{\nu}$, and the scalar curvature
$R$ also have their corresponding scale-covariant expressions: 
\begin{equation}
R^{\nu}_{\mu} = R'^{\nu}_{\mu}  - \kappa^{; \nu}_{\mu}  - \kappa^{ \nu}_{;\mu}
- g^{\nu}_{\mu}\kappa^{ \alpha}_{;\alpha}  -2 \kappa_{\mu} \kappa^{\nu}
+ 2 g^{\nu}_{\mu}\kappa^{ \alpha} \kappa_{ \alpha}  \, .
\label{RC}
\end{equation}
\begin{equation}
R\,=\,R'-6\kappa_{;\alpha}^{\alpha}+6\kappa^{\alpha}\kappa_{\alpha}\,.\label{RRR}
\end{equation}
Here the terms with a prime are the usual expressions in the Riemann
geometry, and the semicolon ` ; ' indicates the usual covariant derivative
with respect to the relevant coordinate.

The main difference with the standard tensor analysis is
that all these expressions contain terms depending on the additional
scalar field through the coefficient of metrical connection $\kappa_{\nu}$
given by the above Equation (\ref{k3}). We notice that the curvature
term has been modified as is the case in the context of the 
``Extended Theories of Gravity'' \citep{Capozziello11}.

\section{The scale-invariant equations for the metric and the gauge fixing}
\label{gauge_condition}

The above developments of the equivalent expressions of the Ricci
tensor and curvature scalar lead to the expression of the general
field equation. We note that it can also be obtained by using the
properties of conformal equations for the Ricci tensor and curvature
scalar, as well as by the application of an action principle, as shown
by \citet{Canu77}. The field equation is 
\begin{eqnarray}
R'_{\mu\nu}-\frac{1}{2}\ g_{\mu\nu}R' -\kappa_{\mu;\nu}-\kappa_{\nu;\mu}-2\kappa_{\mu}\kappa_{\nu}
+2g_{\mu\nu}\kappa_{;\alpha}^{\alpha}-g_{\mu\nu}\kappa^{\alpha}\kappa_{\alpha} \,  = \nonumber \\
=\, -8\pi GT_{\mu\nu}-\lambda^{2}\Lambda_{\mathrm{E}}\,g_{\mu\nu}\, ,  \label{field}
\end{eqnarray}
where $G$ is the gravitational constant (taken here as a true constant)
and $\Lambda_{\mathrm{E}}$ the Einstein cosmological constant. The
energy-momentum intensor $T_{\mu\nu}$ must be a scale-invariant quantity,
as is the first member of the scale-invariant field equation, {{\emph{i.e.}}
$T_{\mu\nu}=T'_{\mu\nu}$. This requirement has some important implications
for pressures and densities \citep{Canu77}. Expressing this condition,
one has: 
\begin{equation}
(p+\varrho)u_{\mu}u_{\nu}-g_{\mu\nu}p=(p'+\varrho')u'_{\mu}u'_{\nu}-g'_{\mu\nu}p'\,.
\end{equation}
The four-velocities $u'^{\mu}$ and $u'_{\mu}$ transform like a co-vector
of order $\mp1$: 
\begin{eqnarray}
u'^{\mu}&=&\frac{dx^{\mu}}{ds'}=\lambda^{-1}\frac{dx^{\mu}}{ds}=\lambda^{-1}u^{\mu}\,,\nonumber \\
u'_{\mu}&=&g'_{\mu\nu}u'^{\nu}=\lambda^{2}g_{\mu\nu}\lambda^{-1}u^{\nu}=\lambda\,u_{\mu}\,.\label{pl1}
\end{eqnarray}
Thus, from the expression of the energy-momentum in-tensor the transformations
for $p$ and $\rho$ follow: 
\begin{eqnarray}
(p+\varrho)u_{\mu}u_{\nu}-g_{\mu\nu}p=(p'+\varrho')\lambda^{2}u_{\mu}u_{\nu}-\lambda^{2}g_{\mu\nu}p'\,,\label{pl}\\
\Rightarrow\;p=p'\,\lambda^{2}\,\;\mathrm{and}\;\varrho=\varrho'\,\lambda^{2}\,.\label{ro2}
\end{eqnarray}
The pressure and density are therefore not scale-invariant, but are
so-called co-scalars of power $n=-2$. A similar scaling appears for
the term containing the cosmological constant,  
$-\lambda^{2}\Lambda_{\mathrm{E}}\,g_{\mu\nu}$;
thus, $\Lambda_{\mathrm{E}}$ is also a co-tensor of power $n=-2$ since 
$\lambda^{2}\Lambda'_{\mathrm{E}}=\Lambda_{\mathrm{E}}$.
The product $\lambda^{2}\Lambda_{\mathrm{E}}$, as it stands in the
field equation (\ref{field}), is evidently gauge invariant (in-scalar)
and thus we have a consistent scale-invariant field equation containing
a non-zero cosmological constant. The best way to see this is to consider
the in-scalar equations (\ref{field}) with one contra-variant and
one covariant index (upper and lower) that are gauge invariant, thus
requiring $\lambda^{2}\Lambda_{\mathrm{E}}$ to be gauge invariant
as well. Here, it is an opportunity to recall the remark by \citet{Bondi90},
who pointed out that ``Einstein's disenchantment with the cosmological
constant was partially motivated by a desire to preserve scale-invariance
of the empty space Einstein equations''. \\

\noindent The field equation (\ref{field}) is undetermined due to
the gauge symmetry of the equations and the same remark applies to
the resulting differential equations either in cosmology or for the Newton-like
approximation. The same problem appears in General Relativity, where
the under-determinacy of GR is resolved by the choice of coordinates
conditions. Here, one needs to impose some gauging conditions to define
the scale factor $\lambda$. 
The term $\lambda^{2}\Lambda_{\mathrm{E}}$ represents the energy
density of the empty space in the scale-invariant context and we made
the specific hypothesis that \textit{{the properties of the empty
space, at macroscopic scales, are scale-invariant}} \citep{Maeder17a}.
This choice is justified since the usual equation of state for the
vacuum $P_{\mathrm{vac}}=-\varrho_{\mathrm{vac}}$ is precisely the
relationship that is permitting $\varrho_{\mathrm{vac}}$ to remain
constant for an adiabatic expansion or contraction \citep{Carr92}.
At the quantum level, this does not necessarily apply, however in
the same way as one may use Einstein's theory at large-scales, even
if it does not apply at the quantum level, we do consider that the
large-scale empty space is scale-invariant. Under the above key hypothesis
one is left with the following condition for empty ``vacuum'' spacetime
deduced from the field equation (\ref{field}): 
\begin{equation}
\kappa_{\mu;\nu}+\kappa_{\nu;\mu}+2\kappa_{\mu}\kappa_{\nu}
-2g_{\mu\nu}\kappa_{;\alpha}^{\alpha}+g_{\mu\nu}\kappa^{\alpha}\kappa_{\alpha}
=\lambda^{2}\Lambda_{\mathrm{E}}\,g_{\mu\nu}.\label{fcourt}
\end{equation}

Now, with the assumption that $\lambda$ is only a function of $t$, 
required by considerations of space homogeneity and isotropy, one 
obtains the non-zero terms $\kappa_{0}$ and $\dot{\kappa_{0}}$
and with Equation (\ref{k3}) one has $\kappa_{0}\,=\,-\dot{\lambda}/\lambda$,
(dots indicating time derivatives). Thus, the above condition (\ref{fcourt})
leads to \citep{Maeder17a} 
\begin{eqnarray}
\ 3\,\frac{\dot{\lambda}^{2}}{\lambda^{2}}\,=\,\lambda^{2}\,\Lambda_{\mathrm{E}}\,\quad
\mathrm{and}
\quad\frac{\ddot{\lambda}}{\lambda}\,=\,2\,\frac{\dot{\lambda}^{2}}{\lambda^{2}} \label{diffd} \, ,
\end{eqnarray}
\noindent
which show similarity with  de Sitter cosmological model, however here
these equations establish a relation between the scale factor $\lambda$
and $\Lambda_{\mathrm{E}}$, which means that  the scalar field $\lambda$, 
$\kappa_{\nu}$, and the cosmological constant are closely related.
In GR, $\Lambda_{\mathrm{E}}$ and thus
the properties of the empty space are considered to not depend on
the matter content of the Universe. We adopt the same assumption here.
This means that the above relations (\ref{diffd}) are always valid,
whatever the matter content. According to discussions by \citet{Maeder17a}
and \citet{MaedGueor19}, their solution is of the form - the cosmic
time gauge: 
\begin{equation}
\lambda\,=\,\frac{A}{t}\,,\label{l}
\end{equation}
where $A=\,\sqrt{\frac{3}{\Lambda_{\mathrm{E}}}}$ follows from the
first equation. Equations (\ref{diffd}) lead to considerable simplifications
in the cosmological equations derived by \citet{Canu77} leading to
solutions different, but no too far from the $\Lambda$CDM models.
These models compare remarkably well \citep{Maeder17a} with  
cosmological observations  showing  an acceleration of the cosmic expansion.

\section{The radial acceleration relation in the Scale-Invariant Vacuum Theory}
\label{acceleration_relation}

\noindent 
\subsection{The weak-field equation in the Scale-Invariant Vacuum  theory}

\label{weakf}

The equivalent of the Newton equation in the scale-invariant framework
was derived from the weak-field approximation of the geodesic equation
\citep{MBouvier79,Maeder17c}. This geodesic equation follows from
an action principle discussed in \citet{Dirac73}, see also \citet{BouvierM78}.
The equation of motion in spherical coordinates is: 
\begin{equation}
\frac{d^{2}\vec{r}}{dt^{2}}\,=\,-\frac{G\,M}{r^{2}}\,\left(\frac{\vec{r}}{r}\right)
+\,\kappa(t)\,\frac{d\vec{r}}{dt}\,.\label{Nvec}
\end{equation}
From Equations (\ref{k3}) and (\ref{l}), the coefficient of metrical
connection has the simple form $\kappa(t)=1/t$, where in the weak
field approximation $t$ is the cosmic time. In the case of the two-body problem, 
in radial coordinates where the trajectory is viewed as a segment of a circle
with a radius $r(t)$, the correction term is proportional to the tangential
velocity $\vec{\upsilon}=d\vec{r}/dt=\vec{\omega}\times\vec{r}$ where $\vec{\omega}$
is the corresponding angular velocity vector. In this framework, the
radial component of the gravity is due to the usual Newtonian term
and has no contribution from the correction term related to $\kappa(t)$.
The effect of the correction term is only on the tangential component
of the motion. 
As discussed for the 2-body problem \citep{MBouvier79,Maeder17c},
this leads to a secular increase of the orbital radius, the
circular velocity keeping constant.
For a general non-circular motion, (\ref{Nvec}) shows an overall enhancement to the radial motion. 
That is, an outgoing motion gets an extra boost outwards and similar for an in-falling motion which is accelerated inwards.
 The effect is strongly reduced at the current age of the Universe. 
 For the early Universe, however, this enhanced flow results in much faster structure formation \citep{MaedGueor19}.
 
One easy way to see where the extra term in the equation (\ref{Nvec}) is coming from is to
extend the notion of parallel transport as defined by the equation
of the geodesics. It has been proven in \cite{BouvierM78} that the
generalized equation of the geodesics follows from a unique action
functional build from the length function of power (-1). Thus, by
extending the usual equation of the geodesics from the Einstein GR
space to the Weyl integrable space with modified Christoffel symbols
(\ref{eq:co-Chr_sym}) and turning the standard covariant derivative
``$;$" into co-covariant derivative ``$*$" one has
$u^{\nu}u_{;\nu}^{\mu}=0\rightarrow u^{\nu}u_{*\nu}^{\mu}=0.$
Based on the definitions in (\ref{eq:co-Chr_sym}) and (\ref{eq:co-cov_der})
this becomes:
\begin{equation}
u^{\nu}u_{*\nu}^{\mu}=u^{\nu}u_{;\nu}^{\mu}
+u^{\nu}[(g_{\nu\alpha}\kappa^{\mu}-g_{\nu}^{\mu}\kappa_{\alpha}-g_{\alpha}^{\mu}\kappa_{\nu})u^{\alpha}
-n\kappa_{\nu}u^{\mu}] \, = \, 0 \, ,
\end{equation}
\begin{equation}
u^{\nu}u_{*\nu}^{\mu}=u^{\nu}u_{;\nu}^{\mu}+(u\cdot u)\kappa^{\mu}-(2+n)(u\cdot\kappa)u^{\mu} \, = \, 0 \, .
\end{equation}
\noindent
This expression shows that for a co-vector $u^{\mu}$ of power $-1$ ($n=-1$) within the
gauge choice for $\kappa_{\nu}$ (\ref{k3}) with $\lambda(t)$ satisfying
(\ref{diffd}) the usual weak-field geodesic equation will have an
additional term that will look like an external force of the form $u^{0}\kappa_{0}u^{i}$.
That is, $u^{\nu}u_{*\nu}^{i}=0\Rightarrow u^{\nu}u_{;\nu}^{i}=u^{0}\kappa_{0}u^{i}.$
Here $i=1,\,2,\,3$ is indexing the spacial components of the four-vector
$u$. Since the three dimensional velocity can be expressed as 
$\upsilon^{i}=dx^{i}/dt=cdx^{i}/dx^{0}=cu^{i}/u^{0}$,
one is arriving  at (\ref{Nvec}).

In what follows, we will try to estimate the magnitude of the ratio
of the correction term $\kappa(t) \, \upsilon\, $ to the usual Newtonian term:
\begin{equation}
x=\frac{\kappa \,\upsilon r^{2}}{GM} \,.
\label{xx}
\end{equation}
Note that $x\thickapprox0$ corresponds to negligible corrections
to the Newton's equation and thus in this case perturbation methods can be
used, while $x\thickapprox1$ and bigger is the strong corrections
regime. Because $\kappa(t)$ and the Hubble constant $H(t)=\dot{a}/a$
are co-scalars of rank $(-1)$, therefore $H(t)/\kappa(t)$ is in-scalar,
that is independent of the general scale factor $\lambda$. Since
in normalized cosmic time units $t=t_{0}=1$ then one has $\kappa(t_{0})=1$
and $H(t_{0})=2$ in the absence of matter (see the discussion after
eq. 32 in \citet{MaedGueor19}); thus, $H(t)/\kappa(t) \, = \,2$
in the totally empty flat universe.

In general, $\xi=H(t)/\kappa(t)=H(\tau)/\kappa(\tau)=H(t_{0})/\kappa(t_{0})=
2\left(1-\Omega_{\mathrm{m}}^{1/3}\right)/\left(1-\Omega_{\mathrm{m}}\right)$.
For $\Omega_{\mathrm{m}}=0.30,\,0.20,\,0.10,\,0.04$ one gets $\xi=0.944,\,1.038,\,1.191$
and 1.371 respectively, which are all of order 1. This way the ratio
$x$ is:

\begin{equation}
x=\frac{\kappa \upsilon \,r^{2}}{GM}\,=\,\frac{H_{0}}{\xi}\frac{\upsilon \,r^{2}}{GM}\,.\label{eq-def_x}
\end{equation}
By using the expression of the Hubble constant $H_{0}$ via the critical
density $\varrho_{c}=3H_{0}^{2}/(8\pi G)$, along with the mean mass
density $\overline{\varrho}$ of the total mass $M$ within a sphere
with a radius $R$ (defined as usual to be $\overline{\varrho}=3M/(4\pi R^{3})$
) one has: 
\begin{equation}
x\,=\,\frac{\sqrt{8\pi G\varrho_{c}/3}}{\xi}\frac{\upsilon}{G}\left(\frac{3r^{2}}{4\pi R^{3}\overline{\varrho}}\right)
=\frac{\sqrt{2}}{\xi}\left(\frac{\varrho_{c}}{\overline{\varrho}}\frac{\upsilon^{2}r^{2}}{GR^{3}}\right)^{1/2}\left(\frac{3r^{2}}{4\pi R^{3}\overline{\varrho}}\right)^{1/2}\,,
\end{equation}
This expression can be written also in the following form: 
\begin{equation}
x=\,\frac{\sqrt{2}}{\xi}\left(\frac{\varrho_{c}}{\overline{\varrho}}\frac{\upsilon^{2}r}{GM}\right)^{1/2}\left(\frac{r^{3}}{R^{3}}\right)^{1/2}\,.
\end{equation}
When the radius of the sphere $R$ defining the mean mass density
$\overline{\varrho}$ coincides with $r$, which is a consistent choice, one obtains: 
\begin{equation}
x\,=\,\frac{\sqrt{2}}{\xi}\left(\frac{\varrho_{c}}{\overline{\varrho}}\frac{\upsilon^{2}R}{GM}\right)^{1/2}\,
=\,\frac{\sqrt{2}}{\xi}\left(\frac{\varrho_{c}}{\overline{\varrho}}\frac{g_{obs}}{g_{bar}}\right)^{1/2}\,,\label{x1}
\end{equation}
where $g_{\mathrm{obs}}=\upsilon^{2}/r$ is the kinematic rotational acceleration
in circular motions, and $g_{\mathrm{bar}}=GM/r^{2}$ is the acceleration
due to the mass present within the radius considered. The above 
expression (\ref{x1}) only applies to spherical distributions of matter. 

In the standard
Newtonian gravity one usually has $\upsilon^{2}r=GM$ which results in: 
\begin{equation}
x\,=\,\frac{\sqrt{2}}{\xi}\left(\frac{\varrho_{c}}{\overline{\varrho}}\right)^{1/2}\,.
\end{equation}
This is the expression of the $x$-ratio when the deviations from
the Newtonian case are small ($x \ll 1$)  with an average matter
density much bigger than the critical density,  $\overline{\varrho}\gg\varrho_{c}$.
A similar expression was used in \citet{Maeder17c} for the discussion
of clusters of galaxies.

\subsection{The radial acceleration relation (RAR) within the SIV context}  \label{rarsivt}

The radial acceleration relation (RAR) expresses the relation
between quantities $g_{\mathrm{obs}}=\upsilon^{2}/r$ and $g_{\mathrm{bar}}=GM/r^{2}$,
\textit{{i.e.}} between the kinematic rotational acceleration
in circular motions and the acceleration due to the mass present within
the radius considered \citep{McGaugh16, Lelli17}. In the Newtonian context, 
$g_{\mathrm{obs}}$ and $g_{\mathrm{bar}}$ are expected to be equal.

Let us start with Equation (\ref{eq-def_x}) written as:
\begin{equation}
x \, = \, \frac{H_0}{\xi}  \frac{(r \, g_{\mathrm{obs}})^{1/2}}{g_{\mathrm{bar}}}\, .
\label{x0}
\end{equation}
\noindent
Quite generally, the ratio $x$ is a function $x(r, t)$ of space and time.
At a given time, one has for the spatial dependence:
\begin{equation}
x(r) \, \sim \, \frac{(r \, g_{\mathrm{obs}})^{1/2}}{g_{\mathrm{bar}}}\, .
\label{xr}
\end{equation}
\noindent
As a result of the time evolution of the system in the scale-invariant context, 
there is a difference between the two accelerations $g_{\mathrm{obs}}(r,t)$
and $g_{\mathrm{bar}}(r,t)$  at a given location. 
Let us write it as $g_{\mathrm{obs}}=g_{\mathrm{bar}}+\Delta g$.
From the meaning of $x$ encoded in its definition via Equation (\ref{Nvec}),
$x$ is the ratio of the difference between the total acceleration
and $g_{\mathrm{bar}}$ with respect to $g_{\mathrm{bar}}$. 
Since the motions in spiral galaxies are essentially circular, the main
component of the total acceleration is still $g_{\mathrm{obs}}$, the centripetal acceleration.
 Indeed, for the Sun the value of the circular
velocity is 248.5 km s$^{-1}$, the radial component 13.0 km s$^{-1}$
and the ``vertical one'' 7.84 km s$^{-1}$ \citep{Schonrich12}.
Thus, one  can write:
\begin{equation}
x(r)\,\approx\,\frac{\Delta g}{g_{\mathrm{bar}}}\,.
\label{dep1}
\end{equation}
From expressions (\ref{xr}) and (\ref{dep1}), 
one has the relation, 
\begin{equation}
\frac{g_{\mathrm{obs}}-g_{\mathrm{bar}}}{g_{\mathrm{bar}}}\,
\sim\,\frac{(r\,\,g_{\mathrm{obs}})^{1/2}}{g_{\mathrm{bar}}}\,.
\label{dep2}
\end{equation}
\noindent
In order to eliminate   the proportionality factor in Equation (\ref{dep2}),
one can consider  two gravitational systems 1 and 2   with respective baryonic gravities $g_{\mathrm{bar,1}}$ 
and $g_{\mathrm{bar,2}}$, observed  at the same   time $t$ and coordinate $r$, with dynamical gravities 
$g_{\mathrm{obs,1}}$ and $g_{\mathrm{obs,2}}$. The ratio of the relative differences 
$\left(\frac{g_{\mathrm{obs}}-g_{\mathrm{bar}}}{g_{\mathrm{bar}}}\right)$,
should according to (\ref{dep2}) behave like: 
\begin{equation}
\frac{\left(\frac{g_{\mathrm{obs}}-g_{\mathrm{bar}}}{g_{\mathrm{bar}}}\right)_{2}}
{\left(\frac{g_{\mathrm{obs}}-g_{\mathrm{bar}}}{g_{\mathrm{bar}}}\right)_{1}}\,=
\,\left(\frac{g_{\mathrm{obs, 2}}}{ g_{\mathrm{obs, 1}}}\right)^{1/2}\,
\left(\frac{g_{\mathrm{bar,1}}}{g_{\mathrm{bar,2}}}\right)\,.\label{corrg}
\end{equation}
\noindent
The $r$-dependence simplifies when the gravities are considered at the same location in the two systems. 
This means that the two systems differ by the importance of their  central mass and/or by their  density distribution.
This equation can also be written as: 
\begin{equation}
\left(\frac{g_{\mathrm{obs}}}{g_{\mathrm{bar}}}\right)_{2}\,=
\,1+\frac{g_{\mathrm{bar,1}}}{g_{\mathrm{bar,2}}}
\,\left(\frac{g_{\mathrm{obs, 2}}}{ g_{\mathrm{obs, 1}}}\right)^{1/2}\,
\left[\left(\frac{g_{\mathrm{obs}}}{g_{\mathrm{bar}}}\right)_{1}-1\right]\,.\label{cocorr}
\end{equation}
This expression relates the values of a particular gravitational system 2
to those of another system 1, chosen as a  reference  at the same time $t$ and location $r$. 
The nature of this reference is discussed below.

We now move to notations often used in the MOND context \citep{Milgrom19},
there $g_{\mathrm{obs}}$ is just $g$ and $g_{\mathrm{bar}}$
is the Newtonian gravity $g_{\mathrm{N}}$. We can write Equation
(\ref{cocorr}) in a compact form:
\begin{eqnarray}
\frac{g}{g_{\mathrm{N}}}&=&1+\left(\frac{g^{1/2}}{g_{\mathrm{N}}}\right)\,k_{1}\,(k_{2}-1)\,,\,\label{gk} \\
&&\mathrm{with}\;k_{1}=\left(\frac{g_{\mathrm{N}}}{g^{1/2}}\right)_{1},\;
\mathrm{and}\;k_{2}=\left(\frac{g}{g_{\mathrm{N}}}\right)_{1}.\nonumber
\end{eqnarray}
\noindent
which is:
\begin{equation}
\frac{g}{g_{\mathrm{N}}}=1+k\left(\frac{g^{1/2}}{g_{\mathrm{N}}}\right)\, ,
\quad\mathrm{with}\,\;k=k_{1}(k_{2}-1)=\left(\frac{g-g_{N}}{\sqrt{g}}\right)_{1}\,.\label{gn} 
\end{equation}
\noindent
We see that the $k$-term only depends on the reference system 1 and 
that $(k_{2}-1)$ cannot be equal to  strictly zero since in any dynamical
situation the additional acceleration term in Equation (\ref{Nvec})
is always present and therefore $k_{2}>1$, thus, $k$ is also positive.
Equation (\ref{gn}) is a second-degree polynomial equation, 
\begin{equation}
g^{2}-g(2\,g_{\mathrm{N}}+k^{2})+g_{\mathrm{N}}^{2}\,=\,0\,,\label{gcarre}
\end{equation}
with two solutions: 
\begin{equation}
g\,=\,g_{\mathrm{N}}+\frac{k^{2}}{2}\pm\frac{1}{2}\sqrt{4g_{\mathrm{N}}k^{2}+k^{4}}\,.\label{sol}
\end{equation}
\noindent
Equation (\ref{gcarre}) can be derived directly from (\ref{dep2}). 
In this case (\ref{x0}) will provide the following expression for $k^2\approx H_0^2 R$. 
We will come back to this expression in our estimate of $k^2$ as discussed in (\ref{k2fin}).
The advantage of using  (\ref{corrg}) is the obvious scale invariance of the expression.
In this respect there are two scale-invariant solutions as seen in (\ref{sol}).
The sign ``+'' should be chosen since $g$ is generally found larger than $g_{\mathrm{N}}$.

There are two limiting cases predicted by this equation.
First, we have the case where the
Newtonian gravity most largely dominates over the effects due to scale
invariance,
\begin{equation}
g_{\mathrm{N}}\,\gg\,\,k^{2} \, ,\quad\mathrm{then}\;\;g\,\rightarrow\,g_{\mathrm{N}}\,,
\end{equation}
\noindent
the dynamical gravity   $\upsilon^2/r$ tends towards the baryonic gravity  $GM/r^2$.
Indeed, as the equations (\ref{corrg}) and the following ones
  concern  the relations of $g_{\mathrm{obs}}$ with $g_{\mathrm{bar}}$ for   given values of the radii,
  this means that very large values  of $g_{\mathrm{bar}}$ are  due to very large masses.
  Thus,  this case occurs for  a very large central mass and/or a high internal density distribution.
Notice that expression (\ref{sol}) may be simplified for gravities $g_{\mathrm{N}}$ much larger than $k^2$.
There,  Equation (\ref{sol})  can be approximated by 
\begin{equation}
g\,\rightarrow\,g_{\mathrm{N}}\,+\,\sqrt{g_{\mathrm{N}}\,k^{2}}\,.
\label{SM}
\end{equation}
This approximation is valid when $g_{\mathrm{N}}$ is at least about two orders of magnitude larger than $k^2$.

The second interesting case occurs when  $g_{\mathrm{N}}$ tends towards zero. This may occur at any value of $r$ due
 to a vanishing central mass or to an extremely faint density distribution. In this case, 
 the dynamical  gravity $g$ tends towards a limiting value $k^2$, 
\begin{equation}
g_{\mathrm{N}}\,\rightarrow \, 0 \, ,\quad\mathrm{then}\;g\;\,\rightarrow\,k^{2}\,.
\label{asy}
\end{equation}
\noindent
Thus, $k^2$ appears as a kind of residual background acceleration in the Universe, in the absence of significant local gravity.
 As the value  of $k^2$ is obtained everywhere for 
the same limit $g_{\mathrm{N}}\,\rightarrow \, 0 $, we may consider that $k^2$ is the same everywhere. 
The well-defined RAR by \citet{Lelli17}, which concerns measurement points for different radii and inner masses, 
also  indicates that a change of  Newtonian gravity, due to mass or distance, results in similar effects.
 
Finally, we recall that we have considered galaxies at the same time, e.g. the present time, this is correct since
the sample studied \citep{Lelli17} is formed of  local galaxies. The value
of the limiting constant $k^2$ needs to be estimated, we  propose to determine it from observations as discussed
in the next section. 

\subsection{An alternative demonstration of the basic equation}  \label{altern}

An alternative way to arrive at the relationship (\ref{xr}) in a more general 
setting can be obtained by the following reasoning. 
As it is well known from kinematics, the total acceleration of a massive object  
can be decomposed in tangential $a_{\mathrm{t}}$ acceleration along 
the trajectory of the object and a normal  $a_{\mathrm{n}}=\upsilon^2/r$ acceleration perpendicular 
to the instantaneous trajectory segment that can be viewed as part of an
arc from a circle of radius $r$. Considering that the SIV equation (\ref{Nvec}) 
predicts an extra acceleration that is along the trajectory of the object, 
then the following expression can be written:
\begin{equation}
\vec{g}=a_{\mathrm{n}}\vec{e_{\mathrm{n}}}+a_{\mathrm{t}}\vec{e_{\mathrm{t}}}=
\frac{\upsilon^{2}}{r}\vec{e_{\mathrm{n}}}+\frac{\upsilon}{t}\vec{e_{\mathrm{t}}},\label{g-kinematics}
\end{equation}
\noindent
here $\vec{e_{\mathrm{n}}}$ and $\vec{e_{\mathrm{t}}}$ are the corresponding unit co-moving directional vectors.
From Newtonian dynamics point of view the total acceleration of a massive object  in a gravitational field would be:
\begin{equation}
\vec{g}=g_{\mathrm{N}}\vec{e_{\mathrm{R}}}+g_{X}\vec{e_{X}}=\frac{GM}{R^{2}}\vec{e_{\mathrm{R}}}
+g_{X}\vec{e_{X}}\thickapprox\frac{G(M+M_{X})}{R^{2}}\vec{e_{\mathrm{R}}},\label{g-dynamics}
\end{equation}
\noindent
here $\vec{e_{\mathrm{R}}}$ is the corresponding unit vector pointing 
from the center of mass of the system to the object under consideration 
that is at a distance $R$, while $g_{X}$ and $\vec{e_{X}}$ represent possible fictitious
acceleration and its direction. In GR as well as in Newtonian mechanics the fictitious accelerations
can be made to be zero upon a suitable choice of coordinates for the observer. 
Within the dark matter paradigm $g_{X}$ is just due to an extra dark matter $M_{X}$ component
while the directions $\vec{e_{\mathrm{R}}}$ and $\vec{e_{X}}$ are expected to coincide.

By looking at the magnitude of the total acceleration $\vec{g}$ via (\ref{g-kinematics}) 
and upon a simple rearrangement of the relevant terms along with the assumption 
$g_{\mathrm{obs}}\thickapprox  a_{\mathrm{n}}$ one can write:
\begin{equation}
x^{2}=\frac{g^{2}}{g_{\mathrm{N}}^{2}}-\left(\frac{a_{\mathrm{n}}}{g_{\mathrm{N}}}\right)^{2}=
\frac{(g-a_{\mathrm{n}})(g+a_{\mathrm{n}})}{g_{\mathrm{N}}^{2}}\thickapprox
\frac{2a_{\mathrm{t}}g_{\mathrm{obs}}}{g_{\mathrm{N}}^{2}} , \label{xat}
\end{equation}
\noindent
%
\noindent
where  $a_{\mathrm{t}}$ is  the magnitude of the tangential acceleration during the motion of a test particle.
The above form leads to  expression  similar to  (\ref{dep2}).
Thus, upon further applying the reasoning from Equation (\ref{dep2})  to  (\ref{gn}), this
 results in similar functional forms like (\ref{gn})  and subsequent expressions, 
\begin{equation}
\frac{g}{g_{\mathrm{N}}}=1+k\left(\frac{g^{1/2}}{g_{\mathrm{N}}}\right)\, ,
\quad\mathrm{with}\,\;k^2=2a_t \, , 
\label{dep3}
\end{equation}
where $k^2$ can be related  
 to $2a_t$ as in  (\ref{xat}) above. In this respect, $k^2$ is model (system)
dependent acceleration; it has something to do with SIV theory as can be seen from its $2a_t$ form;
it also has something to do with the state of the Universe at large cosmic scales, as mentioned above.
Furthermore, notice that (\ref{g-dynamics}) does not seems to result in a relationship 
of the form (\ref{dep3}) because the extra acceleration $g_X$ is co-linear with the baryonic 
acceleration $g_N$, therefore, dark matter models are incapable of 
the universal behavior (\ref{dep3})  as illustrated in what follows.

Notice that the expression in (\ref{dep1}) has been shown to result 
in two identical mathematical expressions (\ref{gn}) and (\ref{dep3}) 
with two potentially different interpretations of the constant $k^2$.
The expression (\ref{gn}) is based on the comparison to a reference system
for the purpose of eliminating the proportionality factor in (\ref{dep2}),
while (\ref{dep3}) does not involve a reference system. 
However, both expressions are about the deviation from the 
Newtonian behavior and show that far from the central mass 
and/or in low-density regime the dynamical acceleration tends towards 
an asymptotic value $k^2$ that demonstrates a non-Newtonian regime.
If this value is to be the same for both mathematical expressions (\ref{gn}) 
and (\ref{dep3}) then one can conclude that 
$\kappa \upsilon \approx (\frac{H_{0}}{\xi})^2 r$,
which is based on (\ref{x0}) and (\ref{xat}),
that is apparently correct due to the definition of $\xi$ and 
the role of $H_{0}$ in relating $\upsilon$ and $r$.
\\

\begin{figure*}
\centering 
\includegraphics[width=14cm,height=9cm]{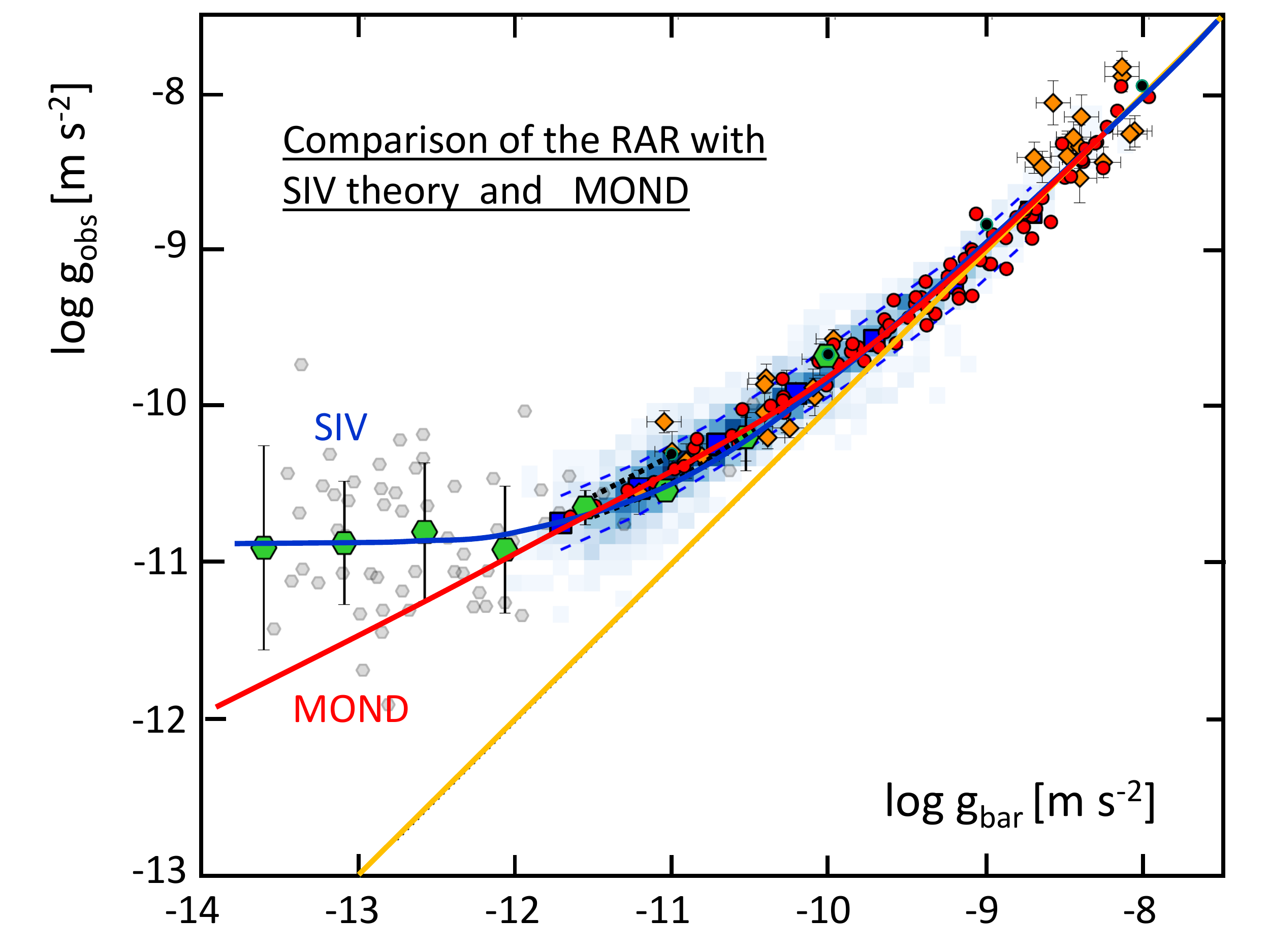} 
\caption{Observed values of $g_{\mathrm{obs}}$ and $g_{\mathrm{bar}}$ for
the 240 galaxies studied by \citet{Lelli17}, forming the radial acceleration
relation (RAR). The big green hexagons represent the binned data of
the dwarf spheroidal galaxies  which satisfy the three quality criteria of
\citet{Lelli17}. The blue curve shows the relation predicted by expression
(\ref{sol}) and given in Table \ref{data1} for SIV with a value of $k^{2}=10^{-10.85}$
m\,s$^{-2}$.  The red curve gives the MOND relation from relation (\ref{mond1}), with $a_{0}=1.20\cdot10^{-10}$
m\,s$^{-2}$, while the orange curve shows the 1:1-line.}
\label{gobs} 
\end{figure*}

In MOND it is assumed that the relation between the gravities
$g$ and $g_{\mathrm{N}}$ is a fundamental law. We do not assume
the same in the scale-invariant theory. In the SIV framework, the fundamental
law is Equation (\ref{Nvec}) obtained in the weak-field Newtonian-like 
approximation of the theory. Relation (\ref{sol}) is its application that is describing 
the connection between $\upsilon^{2}/r$ and $GM/r^{2}$ for the conditions considered. 

\section{Comparison of theories and observations}
\label{Mobs}

\subsection{The radial acceleration relation}
\label{rar}

We examine the radial acceleration relation (RAR) 
studied by \citet{McGaugh16} and \citet{Lelli17} for a sample of
240 galaxies of various morphological types. 
The  relation between $\log g_{\mathrm{obs}}$
and $\log g_{\mathrm{bar}}$ over 6 dex in $g_{\mathrm{bar}}$, from
very massive galaxies to the faint dwarf spheroidals, is shown in
Fig. \ref{gobs}. As stated by Lelli et al., when the baryonic contribution
is measured, the rotation curve follows and reciprocally. For high
gravities, the two gravities are equal, as predicted by the Newton
Law. Below a value $g_{\mathrm{bar}}$ of about $10^{-10}$ m s$^{-2}$,
the RAR significantly deviates from the 1:1 line, $g_{\mathrm{obs}}$
being larger than $g_{\mathrm{bar}}$. These deviations are currently
attributed to dark matter. 

Within the context of the $\Lambda$CDM models of galaxy formation,
the dark-matter description is made in terms of different mass-dependent
density profiles of DM haloes \citep{Dicintio16}. SIV and MOND make sufficiently
specific predictions that are testable against the observational data.
One may also be able to test a large class of Extended Theories of Gravity
\citep{Capozziello11} against the observational data, however, such
models are often funneled into MOND relevant versions 
\citep{2011EPJC...71.1794B}.

We see in Fig. \ref{gobs}, as was also pointed out by \citet{Lelli17},
that the faint dwarf spheroidal galaxies seem to tend on the average
towards a limiting asymptotic value of $\log g_{\mathrm{obs}}$, when
fainter object of lower and lower values of $\log g_{\mathrm{bar}}$
are considered. There is clearly a large scatter in the observations
of the dwarf spheroidals, particularly of the ultra-faint ones that
were recently discovered and where often few stellar velocities are
measured, making the average $g_{\mathrm{obs}}$ uncertain. However,
some of these dwarf galaxies also contain several hundreds or even
thousands of stars measured, for example, as in Fornax, Sextans, and
Sculptor where their dynamics are well studied, see for example \citet{Strigari18}.
In Section \ref{dsph}, the statistics  of the asymptotic
limit of the observed value of $\log g_{\mathrm{obs}}$ is further analyzed.

\begin{table}
\vspace*{0mm}
\caption{Values of $\log g_{\mathrm{obs}}$ (or $\log g$) as a function of
$\log g_{\mathrm{bar}}$ (or $\log g_{\mathrm{N}}$) in the SIV theory from (\ref{sol})
for $k^{2}=10^{-10.85}$, and in MOND from expression (\ref{mond1}) 
with $a_{0}=1.20\cdot10^{-10}$, all accelerations being expressed in m\,s$^{-2}$. }
\label{data1} 
\begin{centering}
{\scriptsize{}}%
\begin{tabular}{c|ccc}
& {\scriptsize{}SIV} & {\scriptsize{}MOND } & \tabularnewline
{\scriptsize{}$\log g_{\mathrm{bar}}$ } & {\scriptsize{}$\log g_{\mathrm{obs}}$ } & {\scriptsize{}$\log g_{\mathrm{obs}}$ } & \tabularnewline 
\hline 
{\scriptsize{}-8.0 } & {\scriptsize{}-7.984} & {\scriptsize{}-8.000} & \tabularnewline
{\scriptsize{}-9.0 } & {\scriptsize{}-8.948} & {\scriptsize{}-8.975} &\tabularnewline
{\scriptsize{}-10.0 } & {\scriptsize{}-9.838} & {\scriptsize{}-9.777} &\tabularnewline
{\scriptsize{}-11.0 } & {\scriptsize{}-10.510} & {\scriptsize{}-10.399} & \tabularnewline
{\scriptsize{}-12.0 } & {\scriptsize{}-10.794 } & {\scriptsize{}-10.941}& \tabularnewline
{\scriptsize{}-13.0 } & {\scriptsize{}-10.844} & {\scriptsize{}-11.454} &\tabularnewline
{\scriptsize{}-14.0 } & {\scriptsize{}-10.849 } & {\scriptsize{}-11.958} &\tabularnewline
\hline 
\end{tabular}
\par\end{centering}{\scriptsize \par}
{\scriptsize{}\vspace{-1mm}
 }{\scriptsize \par}
\end{table}

Equation (\ref{sol}) also predicts a constant limiting value $k^{2}$
for the very low gravities, as shown by its limit (\ref{asy}). We
propose to identify this constant $k^{2}$ with the observed limiting
value of $g_{\mathrm{obs}}$ in the RAR given by \citet{Lelli17}.
Fig. \ref{gobs} suggests a value of $\log g_{\mathrm{obs}}$ of about
-10.85 for this limit, $g_{\mathrm{obs}}$ being expressed in m s$^{-2}$.
This corresponds to a value of $k^{2}=1.41 \cdot10^{-11}$
m\,s$^{-2}$. With this value of $k^{2}$, Equation (\ref{sol})
is fully determined. Table \ref{data1} shows the expected values of $\log g$
for different values of $\log g_{\mathrm{N}}$ according to this equation,
the corresponding values of MOND are also given according to the data
indicated in the caption.

Fig. \ref{gobs} compares the relation in Table \ref{data1} with the values
of $\log g_{\mathrm{obs}}$ and $\log g_{\mathrm{bar}}$ for the 240
galaxies studied by \citet{Lelli17}. The SIV predictions remarkably
agree with observations on the whole range of 6 dex in the accelerations
and also produces the asymptotic behavior observed for dwarf spheroidal
galaxies. The dynamical predictions of the scale-invariant theory,
as expressed by the modified equation of mechanics (\ref{Nvec}),
provides an account of the apparent excess of stellar velocities in
galaxies, with respect to the mass present in them. The excellent
agreement also concerns the spheroidal galaxies as analyzed by \citet{Lelli17},
which describes the RAR over such a large range of accelerations. This
is quite interesting since these are the objects where the ratio
of dark matter to baryonic matter is the highest, reaching up a factor
1000 \citep{Sancisi04}. Further analysis of the spheroidals is
made in Section \ref{dsph} below. For now, the general agreement
found here is one more indication in favor of the scale-invariant
theory, also supported by the dynamics of cluster of galaxies, the
flat galactic rotation curve, the absence of the flat curve in high
redshift galaxies, the growth of the stellar velocity dispersions with ages
\citep{Maeder17c}, as well as the growth of the density fluctuations
in the early Universe \citep{MaedGueor19}, and half a dozen of basic
cosmological tests \citep{Maeder17a}.

\subsection{The asymptotic limit of the dwarf spheroidals}
\label{dsph}

\begin{figure}
\includegraphics[width=9.0cm,height=7.5cm]{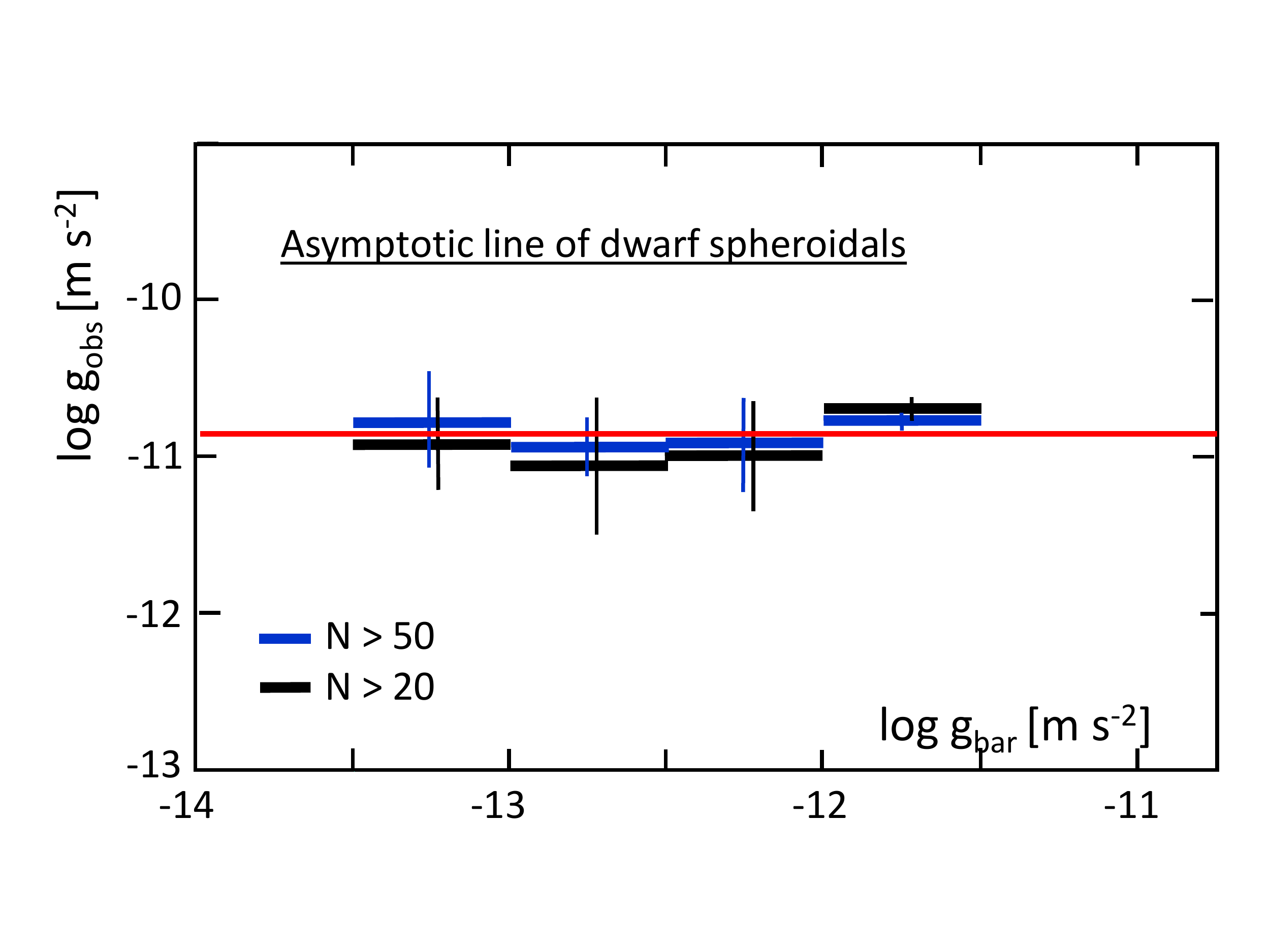} 
\caption{The mean value of $\log g_{\mathrm{obs}}$ per interval of $\log g_{\mathrm{bar}}$
for dwarf spheroidals. The horizontal red line represents the observed value of the asymptotic limit of the dwarf spheroidal galaxies
in Fig. 12 by \citet{Lelli17}.
The thick blue horizontal lines represent the
mean over the interval of $\log g_{\mathrm{bar}}$ considered for galaxies
with a number of observed stars $N\geq50$, in black the average for
galaxies with $N\geq20$ are shown. The thin vertical bars shows the dispersions. }
\label{disp} 
\end{figure}

The asymptotic limit of dwarf spheroidal galaxies found by \citet{Lelli17}
needs to be further examined. A list of 62 dwarf spheroidals in the
Local Group has been given by these authors, with luminosities, 
half-light radii, ellipticities, mean velocity dispersions, number of stars
used to estimate the velocity dispersion, two gravity parameters 
$\log g_{\mathrm{bar}}$ and $\log g_{\mathrm{obs}}$ 
(this last one being estimated on the basis of the velocity dispersion). 
As mentioned by these authors, robust estimates of 
$g_{\mathrm{obs}}$ in dwarf spheroidals (dSphs) can only be
made near the half-light radius, where the effects of anisotropy are
small, however, there is an important exception in the case of the
Fornax and Sculptor dwarf galaxies, which count respectively 2483
and 1365 stars measured (see below). 
Lelli et al. consider three quality criteria to retain or reject dSphs: 
\begin{enumerate}
\item An ellipticity smaller than 0.45. 
\item Limited tidal effects from the host galaxy. 
\item The number $N$ of stars with velocity measurements in a given galaxy. 
They retain only galaxies with $N>8$. 
\end{enumerate}
The mean values obtained by Lelli et al. are represented by the green
hexagons in Fig. \ref{gobs} and define a flat asymptote at about
$\log g_{\mathrm{obs}}=-10.85$.

As the above number $N$ is rather small, we now consider dwarf spheroidals
that have a number $N$ of measurements equal or larger than 20,
and as another option those with $N\geq50$. There are 38 dSphs of
the first group and 16 of the second in the domain considered of $\log g_{\mathrm{bar}}\leq-11.50$.
Fig. \ref{disp} shows the mean values in four intervals, with the
error bars. For values of $g_{\mathrm{bar}}$ varying by a factor
of 100, the value of $g_{\mathrm{obs}}$ remains constant. These two
different samplings of spheroidals with a high number of stars measured
confirm the results of \citet{Lelli17}, with the clear conclusion
that whatever the sampling, the value of $\log g_{\mathrm{obs}}$
is about constant over an interval of 2 dex below about $\log g_{\mathrm{bar}}=-11.50$.

In this context, where  several uncertainties are remaining,  it is worth considering another  test.
The very rich dwarf galaxies Fornax and Sculptor show two chemo-dynamically
distinct stellar components. They both have metal-poor (MP) and metal-rich
(MR) stars,  the MP stars (likely the older ones)
being more broadly distributed than the more concentrated MR stars
\citep{Tolstoy04,Battaglia06}. In these two galaxies, the MP and
MR stars define two different half radii and the velocity dispersions
of the two subgroups are different, nevertheless the values of $g_{\mathrm{obs}}$
are the same. As noted by \citet{Lelli17}: ``Velocity dispersions
and half-light radii seem to conspire to give a constant $g_{\mathrm{obs}}$,
in line with the apparent flattening of the relation at low $g_{\mathrm{bar}}$''.
On the whole, the existence of the asymptotic flat line defined by
the dwarf spheroidals appears as well supported.

\subsection{On the significance and value of the acceleration term $k^{2}$} \label{signif}

From Equations (\ref{gk}) and (\ref{gn}), the parameter $k$ may
be expressed in the following form: 
\begin{equation}
k\,=\,k_{1}(k_{2}-1)\,=\,\left(\frac{g_{\mathrm{N}}}{g^{1/2}}\left(\frac{g}{g_{\mathrm{N}}}-1\right)\right)_{1}\,
=\,\left(\frac{g-g_{\mathrm{N}}}{g^{1/2}}\right)_{1}\,  .\label{valk}
\end{equation}
The above relation  is evidently consistent with the definition given by (\ref{gn}). It also  shows that if $g_{\mathrm{N}}$
tends towards zero, then the dynamical gravity $g$ would tend to $k^2$. Thus, 
$k^2$ appears as a background limiting  value of the dynamical gravity for vanishing Newtonian gravity,
independently of the radius considered. The origin,  in the Newtonian approximation of the scale-invariant theory,
of a non-vanishing dynamical  gravity  $k^2$, arises from the additional acceleration term in Equation (\ref{Nvec}). 
The accumulated effect of this term over the ages, with the assumption that the galaxies have the same age, 
leads to the resulting acceleration term. Physically, in the very faint systems  (with $g_{\mathrm{N}} \, \rightarrow \,0 $), 
this background dynamical acceleration manifests itself as a velocity dispersion becoming larger at larger radii  
(leading to a constant acceleration), a behavior that is  illustrated by the two chemically different 
components of the dwarfs Fornax and Sculptor.

Thus, the value of the background  acceleration, $k^2$ is logically related to some cosmological properties of the Universe,
in particular to the average matter density and its fluctuations. A detailed estimate  is beyond the 
scope of the present paper, however, we may try to estimate the order of magnitude of $k^2$.
The dynamical gravity is always larger or equal to the Newtonian gravity, thus when 
$g_{\mathrm{N}}\rightarrow  0$, we may expect that,
\begin{equation}
k^{2}\,  \rightarrow min[g_{\mathrm{N}}]\,=\,min\left[\frac{GM}{R^{2}}\right]\,
=\,min\left[\frac{4\pi}{3}G\,\overline{\varrho}\,R\right]\, .
\label{kg}
\end{equation}
\noindent
For the density, we take the mean density of the Universe 
$\rho_{\mathrm{m}}=\Omega_{\mathrm{m}}\,\rho_{\mathrm{c}}$,
with the critical density $\rho_{\mathrm{c}}=3\,H_{0}^{2}/(8\pi G)$.
Notice that these expressions result in $min[g_{\mathrm{N}}]\approx H_{0}^{2} R$ 
that is also the mathematical relation for $k^2$ 
as discussed earlier in the paragraph after Eq. (\ref{sol}).
For $R$, some fraction $f$ of the Hubble scale $c/H_{0}$ is reasonable.
Thus, one obtains the following estimate: 
\begin{equation}
k^{2}\,\approx
\,\frac{1}{2}\,f \, \Omega_{\mathrm{m}}\,c\,H_{0}\,.
\label{k2fin}
\end{equation}
\noindent
Thus, the value of the minimum average acceleration $k^2$ is logically related to some cosmological properties of the Universe,
in  particular to the average matter density.
Interestingly enough,  
$k^{2}$ depends on the constant $c$ multiplied by the present expansion
rate $H_{0}$, a dependence also present
in the acceleration term $a_0$ of MOND \citep{Milgrom09,Milgrom15}. 
Here  in (\ref{k2fin}), there is a factor $f$ necessarily smaller than one,   which thus leads to a value of $k^2$ lower 
than the MOND parameter $a_0$, because the equations are also  different. Furthermore,
the  above expression (\ref{k2fin}) implies that the minimum average dynamical acceleration would
vanish only for an empty Universe.

The result above is consistent with a simple dimensional argument based on Equations (\ref{xat}) and (\ref{dep3}) which
shows that  $k^2$ can be linked to  the tangential acceleration $a_{\mathrm{t}}=\upsilon/t$ 
in the case of a specific object on cosmological scales.
An upper-value estimate of $k^2$ based on $a_{\mathrm{t}}$ 
as linked to the large scale cosmology parameters results in:
\begin{equation}
 k^2   \sim  a_{\mathrm{t}}=\upsilon/t= (H_0 \, R_{\mathrm{lim}})/t_{0}  =   (H_0 \, c\,t_0)/t_0=  H_0 \, c \,,
 \label{hoc}
 \end{equation}
 \noindent
where $t_0$ is the age of the Universe.
 This way, we may understand the general dependence on $c\,H_0$, 
while the above estimate (\ref{k2fin}) contains a factor $f$ smaller than 1 
and demonstrates  the influence of the total mass of the Universe within the past-causal cone 
of an arbitrary point that  accounts for the matter density via the term $\Omega_{\mathrm{m}}$.	

Let us examine  numerically  the approximation
(\ref{k2fin}). We  take a value of $H_{0}=70$ km s$^{-1}$ Mpc$^{-1}$,
thus $c\,H_{0}=6.801\cdot10^{-10}$ m s$^{2}$. For 
$\Omega_{\mathrm{m}}=  0.20$ or 0.30,  we get 
$\frac{1}{2} \Omega_{\mathrm{m}}\,c\,H_{0}= 6.80$  or  $10.02$ in units $10^{-11}$ m\,s$^{-2}$ respectively. 
In Section \ref{rar}, we considered the fact   that $k^2$ could correspond to the asymptotic 
limit of the RAR with a value of $k^{2}=1.41 \cdot10^{-11}$ m\,s$^{-2}$. 
This means that the  numerical factor $ f \approx  0.21$ or 0.14 respectively.
This factor is evidently different, but not so much, from the ratio  (0.18) of the  
MOND constant $a_0 = 1.20 \cdot10^{-10}$ m s$^{2}$
to the product $c \, H_0$, since the equations  are  different. 
Nevertheless,   as shown in the next Section, over a large range 
of gravities the two different approaches 
 give the same numerical  results.

Within the dark matter paradigm (\ref{g-dynamics}) the value of $k^{2}$,
according to (\ref{valk}), is expected to be
$k^{2}=(g-g_{\mathrm{N}})^2/g=(GM_{X}/R^2) (M_{X}/(M+M_X))$ 
and in the limit $R\rightarrow R_{\mathrm{lim}}$ it becomes
$k^{2}=(g-g_{\mathrm{N}})^2/g\rightarrow G M_{X}/R_{\mathrm{lim}}^2 \times
 (\Omega_{\mathrm{DM}}/(\Omega_{\mathrm{b}}+\Omega_{\mathrm{DM}}))$,
thus the factor $\Omega_{\mathrm{m}}$ in (\ref{k2fin}) is replaced
by $\Omega^2_{\mathrm{DM}}/(\Omega_{\mathrm{b}}+\Omega_{\mathrm{DM}}))$
which is a dimensionless factor of  
$\Omega^2_{\mathrm{DM}}/\Omega^2_{\mathrm{m}}$ compared to 
$\Omega_{\mathrm{m}}=\Omega_{\mathrm{b}}+\Omega_{\mathrm{DM}}$. 
For $\Lambda$CDM baryonic matter fraction 
$\Omega_{\mathrm{b}}=4\%$ and $\Omega_{\mathrm{DM}}=25\%$ 
this will result in an factor of about $(25\%/29\%)^2=0.74$ which is  $\approx5$
times bigger than the SIV value of $k^2$ given by (\ref{k2fin}). 
Furthermore, notice that (\ref{g-dynamics}) predicts 
$\log g - \log g_{\mathrm{N}}= \log (M_{\mathrm{tot}}/M_{\mathrm{bar}})\approx \log 7 \approx 0.8$ 
that is a factor of 4 times smaller than what is seen in Fig. \ref{MONDSC},
which is most likely due to the fact that dark matter (\ref{g-dynamics}) 
cannot result in a relationship of the form (\ref{dep3}).

\subsection{Comparison and discussion}   \label{Comp}

\begin{figure*}
\centering 
\includegraphics[width=0.75\textwidth]{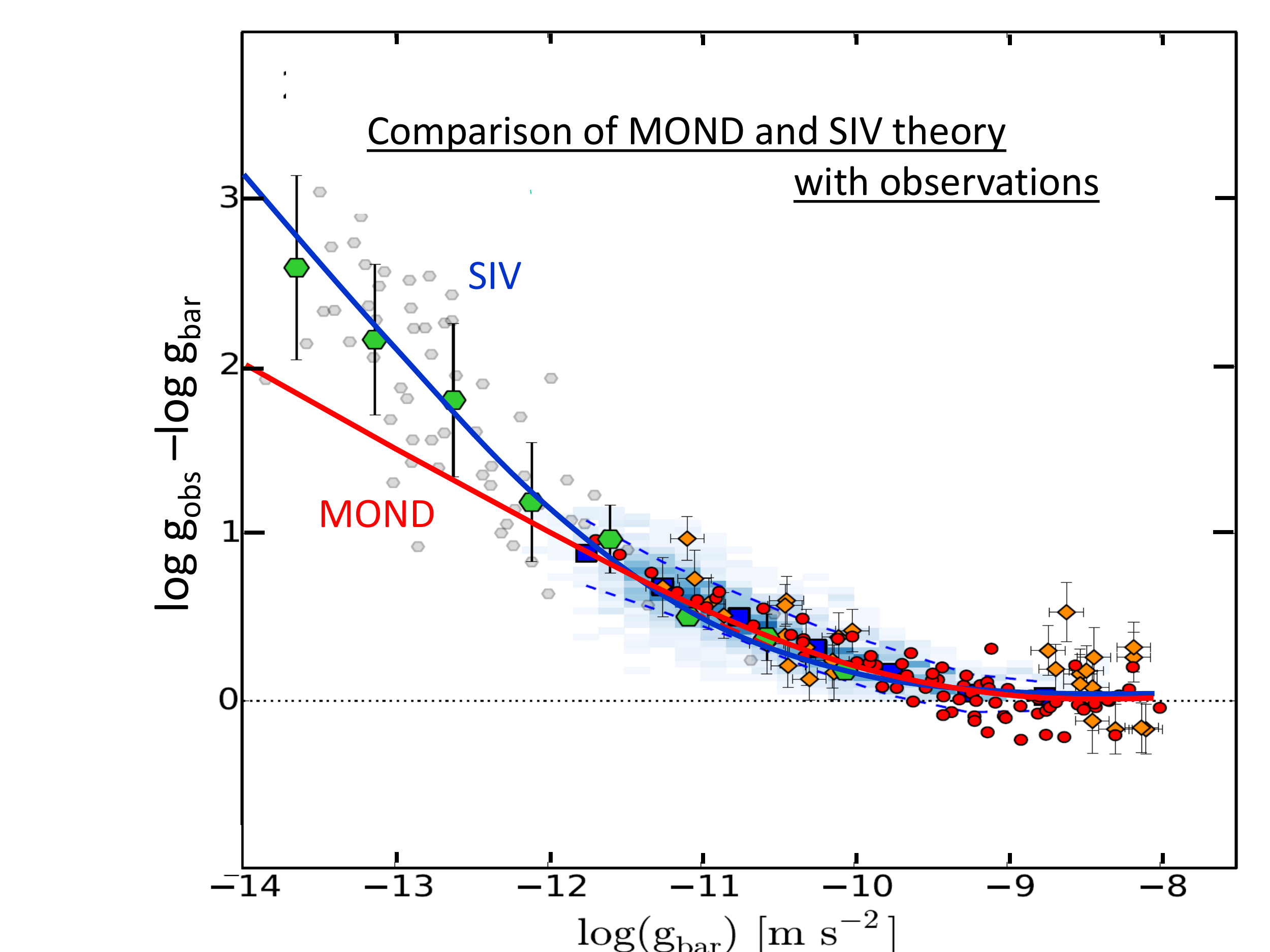}
\caption{The observed deviations $\Delta\log g=(\log g-\log g_{\mathrm{N}})$
for the 240 galaxies analyzed by \citet{Lelli17}. The quantity $\Delta\log g$
is also equal to $\log(\frac{M_{\mathrm{tot}}}{M_{\mathrm{bar}}})$,
the $\log$ of the ratio of the total mass (including the supposed
dark matter) to the baryonic matter. The MOND and SIV model predictions
are compared with the observations. }
\label{MONDSC} 
\end{figure*}

First, on the theoretical side, MOND is based on an hypothesis of scale invariance \citep{Milgrom09},
which assumes an invariance to space and time dilatations with a scale factor $\lambda$ independent on time.
This is a different and less general invariance than the SIV hypothesis,
which assumes space-time invariance, with a scale factor $\lambda(t)$  dependent on time
(see Sect. \ref{base}). Nevertheless, there is some proximity between
the two. For example, in MOND, the orbital velocity around a bounded
mass M becomes independent of the size of the orbit \citep{Milgrom14}.
In SIV context, the circular velocity keeps constant during the secular increase
of the orbital radius of a moving particle.

There is however an important question about MOND, which assumes a relation  $a_0  \sim  c\, H_0$
between the so-called ``MOND fundamental constant'' $a_0$  and the Hubble constant $H_0$. This 
raises the question of what happens to this relation at other epochs.  We must logically wonder whether
these ''constants'' do not depend on time as is the case for the expansion rate $H(z)$.  If this is the case, this would clearly 
favor a variable scalar field, with a variable scale factor $\lambda(t)$, as assumed in the scale-invariant theory.

Keeping the previous notations, the expression of the acceleration
of gravity in the MOND theory may be written as \citep{MilgromS08,Milgrom16},
\begin{equation}
g\,=\,\frac{g_{\mathrm{N}}}{\left[1-e^{-(g_{\mathrm{N}}/a_{0})^{\frac{1}{2}}}\right]}\,,\label{mond1}
\end{equation}
an expression also used for example by \citet{McGaugh16} and \citet{Dutton19}.
There, the term $a_{0}$ represents a constant acceleration equal
to about $a_{0}=1.20\cdot10^{-10}$ m s$^{-2}$. The MOND theory has
two limits: first, for gravities $g_{\mathrm{N}}\gg a_{0}$, the dynamical
gravity tends towards the Newtonian value $g_{\mathrm{N}}$. 
The second limit is:
\begin{equation}
\mathrm{for} \quad g_{\mathrm{N}}\ll a_{0}\,, \quad g \, \rightarrow  \sqrt{g_{\mathrm{N}}\,a_{0}}\,, 
\label{sq2}
\end{equation}
\noindent
\citep{Milgrom83,Milgrom09, Milgrom13,Milgrom16}.
This is the so-called deep MOND limit for weak gravities. Several
other overall expressions, different from Equation (\ref{mond1}) have been
considered over the years and they have been closely compared by \citet{Dutton19}.
As one can see from (\ref{sq2}), the MOND asymptotic limit of $g$ related to 
vanishing baryonic gravity tends to zero. Thus, when $g_N$ tends to zero so is $g$.
In this respect, other RAR studies, such as that by \citet{DiPaolo19}, do not show the non-zero asymptotic limit
of $g$ when $g_N$ tends to zero. In this case,  a critical point  may be  that   the stellar mass distribution "is estimated
kinematically by means of mass modelling of the rotation curve".  An explicit connexion between the velocity,
the baryonic fraction and $g_{\mathrm{bar}}$ is used (see their Equation 7), while the $g_{\mathrm{bar}}$ estimates
should be  independent of velocity properties.


In the line of the mentioned proximity between MOND and the present work,
it is interesting to compare the  above  low gravity limit in MOND as given
by Equation (\ref{sq2})  with the corresponding limit obtained in the SIV Equation (\ref{SM}).
In both cases, there is a dependence on the square root of a product of $g_{\mathrm{N}}$
and a constant term. In  MOND, this applies to the expression of $g$ while in the SIV theory
this applies to the difference $g-g_{\mathrm{N}}$. The constants are evidently 
different since the expressions are different, but the overall curves 
are very similar  for  about $\log g_{\mathrm{N}}   >-11$, as illustrated by Figs. \ref{gobs} and \ref{MONDSC}.\\

\noindent 
Let us estimate the differences $\Delta\log g=\log g-\log g_{\mathrm{N}}$
as a function of $\log g_{\mathrm{N}}$ for the two theories. Quantities
$\Delta\log g$ represent the vertical deviations from the 1:1 line
in Fig. \ref{gobs}. These deviations also correspond to $\log(\frac{M_{\mathrm{tot}}}{M_{\mathrm{bar}}})$,
the log of the ratio of the total mass (including dark matter) to
the baryons mass \citep{Lelli17}. The values $\Delta\log g$ for
the scale-invariant theory are based on the data of Table \ref{data1}, for MOND
they are obtained from Equation (\ref{mond1}). These data are summarized
in Table \ref{data2}. The differences between the two remain very small at high
accelerations, and thus the two theories are in excellent agreement down
to $\log g_{\mathrm{N}}=-11.50$. However, for $\log g_{\mathrm{N}}\leq-12$,
the differences become significant, the values of $\Delta\log g$
predicted by SIV  theory becoming much larger than for MOND. This a consequence
of the horizontal asymptote in Fig. \ref{gobs}, while at
the lowest gravities the MOND curve continues to go down, as predicted
by Equation (\ref{mond1}) \citep{Milgrom16}.

\begin{table}
\caption{Values of differences between $\log g$ and $\log g_{\mathrm{N}}$
for the scale-invariant vacuum theory and for MOND. The values for
SIV are derived from Table \ref{data1} based on Equation (\ref{sol}), for
MOND the values are obtained from Equation (\ref{mond1}), with $a_{0}=1.20\cdot10^{-10}$
m\,s$^{-2}$. }
\label{data2}
\begin{centering}
{\scriptsize{}}%
\begin{tabular}{cccc}
{\scriptsize{}$\log g_{\mathrm{bar}}$ } & {\scriptsize{}$\Delta\log g$ } & {\scriptsize{}$\Delta\log g$ } & \tabularnewline
 & {\scriptsize{}SIV } & {\scriptsize{}MOND } & \tabularnewline
\hline 
 &  &  & \tabularnewline
{\scriptsize{}-8.0 } & {\scriptsize{}0.016} & {\scriptsize{}0.000 } & \tabularnewline
{\scriptsize{}-9.0 } & {\scriptsize{}0.052} & {\scriptsize{}0.025 } & \tabularnewline
{\scriptsize{}-10.0 } & {\scriptsize{}0.162} & {\scriptsize{}0.223 } & \tabularnewline
{\scriptsize{}-11.0 } & {\scriptsize{}0.490} & {\scriptsize{}0.601 } & \tabularnewline
{\scriptsize{}-12.0 } & {\scriptsize{}1.206} & {\scriptsize{}1.059 } & \tabularnewline
{\scriptsize{}-13.0 } & {\scriptsize{}2.156} & {\scriptsize{}1.546 } & \tabularnewline
{\scriptsize{}-14.0 } & {\scriptsize{}3.151} & {\scriptsize{}2.042 } & \tabularnewline
\hline 
\end{tabular}
\par\end{centering}{\scriptsize \par}
{\scriptsize{}\vspace{-1mm}
 }{\scriptsize \par}
\end{table}

The MOND theory can be recovered from the ``$f(R)$-theories of Gravity''  by \citet{Capozziello06},
already mentionned in Section \ref{base}.  However,  the $f(R)$ theories are
 more general than MOND, see also \citep{Borka16,Capozziello17}. According to these authors, 
several  issues of fundamental physics suggest that higher order terms must enter the gravity Lagrangian.
Thus the action writes
\begin{equation}
A = \int d^4x \sqrt{-g} \,  [f(R) + \mathcal{L}_m] \, ,
\end{equation}
where $f(R)$ is some function of the Ricci curvature scalar.  Changing the gravity Lagrangian 
modifies the potential in the weak field approximation. The case $f(R)=R$ corresponds to GR.  
A  power law  $f(R) =f_0 \, R^n$ is often considered, with $f_0$ being some numerical constant. In this case, the resulting potential $\Phi$ 
 takes a simple form, in which  appears  an additional  fundamental radius  
   $r_{\mathrm{c}}$ (in addition to the Schwarzschild radius). The corresponding
  velocity  of circular motions becomes
 \begin{equation}
 v^2_{\mathrm{c}}(r) = \frac{GM}{2\,r} \left[1 +(1-\beta) \left(\frac{r}{r_{\mathrm{c}}}\right)^{\beta}\right] \, ,
 \label{vrotC}
 \end{equation}
 where  the parameter $\beta$ may depend on the scale of the objects  considered. A value $n=1$ and $\beta=0$ 
 applies to the Solar System, while from  their  velocities and 
 luminosities, a value $\beta=0.817$ is adopted for galaxies.
 For $\beta$ between 0 and 1, the predicted velocities are larger than the standard values and 
 thus   this ``may fill the gap between theory and observations without the
need of additional dark matter'' \citet{Capozziello06}. These authors show  
an excellent fit of the rotation curve with the  rotation profiles (\ref{vrotC}) for the above parameter $\beta$. 
For a value of $n=3/2$, the $f(R)$-theory
converges to a MOND-like acceleration  $a \simeq (a_o GM)^{1/2} /r$.  There is  also 
a relation between the MOND parameter $a_0$ and  the new
radius $r_{\mathrm{c}}$ according to \citet{Capozziello17}. These authors
 show an excellent agreement of their model with the baryonic
Tully-Fisher relation, as MOND is doing. These developments well illustrate the variety of approaches of the 
Extended Theories of Gravity in order to account for the flat high rotation curves of galaxies

We now  turn to the comparison with the observations. There are
undoubtedly several successes of MOND in galaxy dynamics, making
it a challenging theory to the $\Lambda$CDM models, as emphasized
by \citet{McGaugh15}. The comparison of MOND and SIV predictions is made in
Fig. \ref{MONDSC}. Over the whole range from the large elliptical
galaxies to the low mass spirals down to about $\log g=-11.50$, the agreement
between the two curves, as well as with the observations is excellent.
It is only for extremely low gravities below the above value that
the two theoretical curves disagree. At $\log g_{\mathrm{bar}}=-13.50$,
 which corresponds to the lowest observed values, the ratio of $g/g_{\mathrm{N}}$
(or $M_{\mathrm{tot}}/M_{\mathrm{bar}}$) predicted by MOND is about
59 while that by SIV is 478, very close to the value supported by
the observations. Thus, the MOND predictions and the observations
of the dwarf spheroidals do not fit, a point also mentioned by \citet{Dutton19}.
The dwarf spheroidals, which are the objects where the amount of dark matter with respect
to baryons is the highest, appear to have a critical role in constraining theories. There are
is still uncertainties in the relevant data,  and thus  observations of greater accuracy 
of dwarf spheroidals are very needed.

\section{Conclusions}\label{Concl}
The SIV theory is based on the key hypothesis that the macroscopic empty space
possesses a scale-invariant symmetry with respect to time and space.
Within the Weyl Integrable Geometry, this results in a modified equation
of motion (\ref{Nvec}). This equation leads to a specific relationship
(\ref{dep2}) between the observed kinematical acceleration 
$g_{\mathrm{obs}}=\upsilon^{2}/r$
and the Newtonian acceleration due to baryonic matter 
$g_{\mathrm{bar}}=GM/r^{2}$.
The relationship is further explored in general terms and leads to
the two-point correlation (\ref{cocorr}) which can be cast in the
key expression (\ref{sol}). Alike MOND, SIV theory well accounts for the
RAR down to $\log g_{\mathrm{bar}}=-11.50$ as seen in Fig. \ref{gobs}.
For lower gravities, SIV is still in full agreement, while this is
not the case for MOND, e.g. the case of the dwarf spheroidals
discussed above. Via (\ref{sol}) SIV predictions naturally accounts for the
horizontal asymptotic limit (\ref{asy}) of the RAR relation, which corresponds to a minimum acceleration of about
$\log(g)=-10.85$ (where $g$ is expressed in m $\cdot$ s$^{-2}$).
This empirical value,  which critically depends on the dispersion velocities of dSph galaxies, appears
stable with respect to a reasonable set
of sampling selection choices as demonstrated by Fig. \ref{disp}.
A theoretical interpretation of this constant gives a reasonable
agreement.

On the whole, the above results suggest, in addition to other tests such as the 
growth of the density fluctuations  \citep{MaedGueor19},  that there is no need for 
dark matter and that the RAR and related dynamical properties of galaxies 
can be interpreted by a modification of gravitation, as proposed by the Scale-Invariant Vacuum theory.

\section*{Acknowledgements}
A.M. expresses his deep gratitude to his wife and to D. Gachet for their continuous support.
V.G. is extremely grateful to his wife and daughters for their understanding and family support  
during the various stages of the research presented. This research did not receive any specific 
grant from funding agencies in the public, commercial, or not-for-profit sectors.

\bsp	
\label{lastpage}
\end{document}